\documentclass{xray_symp}
\usepackage{graphics}

\begin{document}
\title{Evolution of the ROSAT AGN Luminosity Function
}
\author{T. Miyaji\inst{1} \and G. Hasinger \inst{1} \and 
M. Schmidt \inst{2}}  
\institute{Astrophisikalisches Institut Potsdam, An der Sternwarte 16,
Potsdam, D-14482, Germany
\and California Institute of Technology, Pasadena, CA 91125, USA}
\titlerunning{ROSAT AGN Luminisity Function}
\maketitle

\begin{abstract}
 We present the results and parameterization of the 0.5-2 keV
Luminosity Function of AGNs from various ROSAT Surveys, ranging
from the ROSAT bright Survey from the ROSAT All-Sky Survey (RASS)
to the Ultra-Deep survey on the Lockman hole. A Luminosity-dependent
density evolution model, where the density evolution rate drops at
low luminosities, gives an excellent parametric description of
the overall XLF covering wide ranges of redshift and luminosity.

 The number density evolution of high-luminosity AGNs in our sample
shows a similar behavior to optical and radio surveys, except that
we do not find evidence for the rapid decrease of the number
density at $z>2.7$. The discrepancy is marginally significant and
including more deep survey results would make better determination
of the behavior. 
 
\end{abstract}

\section{Introduction}

 Strong X-ray emission is a prominent key character of an AGN activity 
(we use the term ``AGNs'' for Seyfert nuclei and QSOs
collectively). Thus unbiased X-ray surveys and identifications
are important for investigating cosmological evolution of AGN 
activities. A combination of ROSAT surveys, ranging from the 
RASS to the ROSAT Ultra Deep Survey on the Lockman Hole, provides a large 
sample of soft X-ray selected AGNs. In this article, we report 
basic results of our work on the soft (0.5-2 keV) X-ray luminosity 
function (SXLF) of AGNs and its evolutionary properties, using a 
combined {\it ROSAT} sample of about 670 AGNs. 

 A construction of a population synthesis model of the  
Cosmic X-ray Background (CXRB) with a combination of 
unabsorbed ``type 1'' and intrinsically absorbed ``type 2'' AGNs
are presented elsewhere (Miyaji et al. 1999; see also
Schmidt et al. 1999). Unless otherwise noted, 
we use $H_0=50\;h_{50}$ $[{\rm km\;s^{-1}\;Mpc^{-2}}]$ 
and $(\Omega_{\rm m},\Omega_\Lambda)$ $=(1.0,0.0)$ in our 
calculations. The symbol $L_{\rm x}$ refers to the luminosity in 
$[{\rm erg\;s^{-1}}]$ in 0.5-2 keV using a K-correction assuming
a power-law photon index of $\Gamma=2$. This is equivalent to
the no K-correction case, and more realisticaly, this should
be considered as the 0.5$(1+z)$ -- 2$(1+z)$ [keV] luminosity
at the source, given the variety of AGN spectra. 
The symbol $S_{\rm x14}$ refers to the 0.5-2 keV flux measured in  
$[10^{-14}{\rm erg\,s^{-1}\,cm^{-2}}]$.

\begin{figure}[b]
\begin{center}
\resizebox{0.8\hsize}{!}{\includegraphics{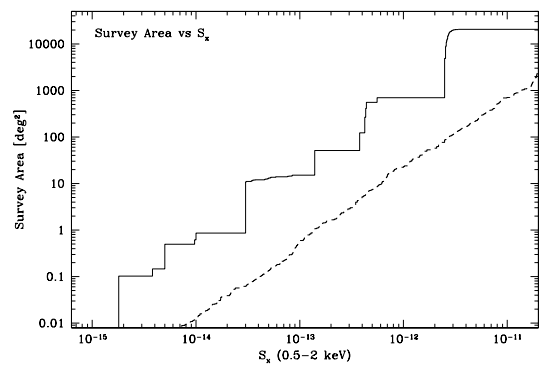}}
\end{center}
\caption[]{The survey area of the combined sample
are plotted as a function of the limiting 0.5-2 keV
flux limit (solid line). For reference, $[N(>S)]^{-1}$ 
for all the X-ray sources is overplotted (dashed-line)
in the same scale.}
\label{fig:area}
\end{figure}

\section{Sample}

 We have constructed a combined sample from various ROSAT surveys
as summarized in Table \ref{tab:surv}. RBS and SA-N are from
RASS, RIXOS is a serendipitous survey and others are 
pointed deep surveys. For the UKD and Marano samples,
we have included sources identified with QSOs and NELGs, but 
used the objects with $S_{\rm x14}\geq 0.5$, in order to minimize 
the possibility of including significant number of misidentified 
sources (see Schmidt et al. 1998). For the Lockman hole, we have
combined the very deep ($\sim 1$ Msec) HRI data (HRI offaxis
$\leq 12\arcmin$ with a limiting flux of 
$S_{\rm x14}^{\rm lim} = 0.19$) and PSPC data 
(HRI offaxis $>12\arcmin$, $S_{\rm x14}^{\rm lim} = 0.38-0.5$ depending 
on the PSPC off-axis angle). For the PSPC data, we have
used only pulse-height channels corresponding 0.5-2 keV and, for 
both the PSPC and HRI data, the countrate-to flux conversion has been 
made using a photon index of $\Gamma=2$ corrected for the effect 
of the absorption in our galaxy. Unlike some previous works,
we have included all objects which have AGN indications (except BL-Lac
objects), not only the ones with apparent broad lines. Some of these 
objects would have been classified as ``Narrow Emission Line Galaxies'' 
(NELGs) by other groups and thus would have been excluded from their 
analysis.

\begin{table}[t]
\caption[]{ROSAT Surveys used in the analysis}
\begin{center}
\begin{tabular}{lccc}
\hline\hline
 Survey$^{\rm a}$ & $S^{\rm lim}_{x14}$ & Area & No. of  \\
       & $[{\rm erg\,s^{-1}\,cm^{-2}}]$ & $[{\rm deg}^2]$ & AGNs  \\
\hline
 RBS   & $\approx 250$ & $\approx 20000$ & 223 \\
 SA-N  & $\approx 13$ & $\approx 640$  & 130 \\
 RIXOS & $3.0$  & $\approx 15$     & 205 \\    
 NEP   & $1.0$  & $\approx 0.2$      & 13  \\
 Marano & $0.5$ & $\approx 0.2$   & 28  \\
 UKD   & $0.5$  & $\approx 0.16$  & 29  \\
 RDS & $0.19-0.5$ & $\approx 0.1$ & 62  \\
\hline
\end{tabular}
\end{center}
\label{tab:surv}
{
$^{\rm a}$ Abbreviations/Reference -- RBS: The ROSAT Bright Survey
(Schwope et al. 1998), SA-N: The Selected Area-North (Appenzeller
et al. 1998, RIXOS: The ROSAT International 
X-ray Optical Survey (Mason et al. 1998), NEP: The North Ecliptic Pole
(Bower et al. 1996)
Marano: The Marano field (Zamorani et al. 1998), 
UKD: The UK Deep Survey (McHardy et al. 1998), RDS: The {\it ROSAT} Deep 
Survey on the Lockman hole 
(Hasinger et al. 1998; Schmidt et al. 1998)
}
\end{table}

 Fig. \ref{fig:area} shows the available area of the combined sample
as a function of the limiting flux with the inverse of the 
AGN $N(>S)$ in our sample. 

\section{Description of the overall SXLF}

 Using the maximum-likelihood method, we have fitted the unbinned
sample  with a number of SXLF models, including the
Pure Luminosity Evolution (PLE), Pure Density Evolution
(PDE), and the Luminosity-dependent density evolution (LDDE)
models. The overall fit has been made for the redshift range
$0.015<z<5$ and $41.7\le {\rm Log}\;L_{\rm x} \le 47.0$. 
We have tested the best-fit models in each class with a two 
dimensional Kologomorov-Smirnov (2D K-S)
test (Fasano \& Franceschini 1987). The details of the statistical 
analysis will be discussed elsewhere (Miyaji et al. 1998b). The basic
results are:

\begin{figure}[b]
\resizebox{\hsize}{!}{\includegraphics{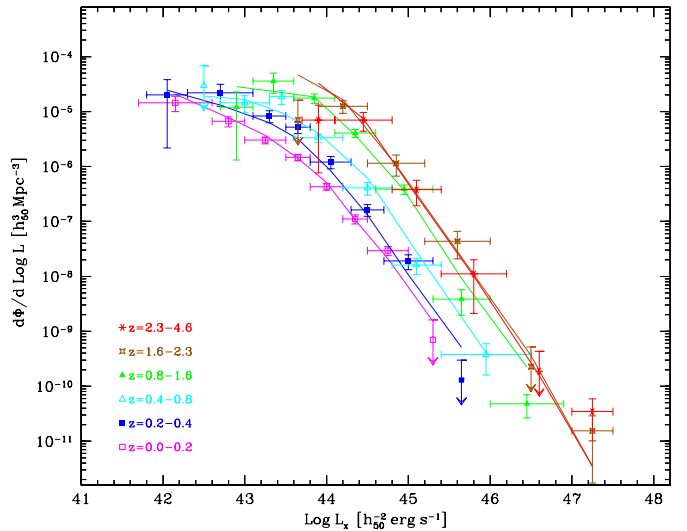}}
\caption[]{The $1/V_a$ estimates of the SXLF in different
redshift bins (as labeled) are shown. One sigma error bars are shown for 
each bin. Important upper limits are shown as downward arrows.
The top tick and symbols mark of the upper limit correspond to 2.3 
objects (90\%) and one object per bin respectively. The solid 
lines show rough predictions from the best-fit LDDE model (see below) 
for the redshift bins.}

\label{fig:xlf}
\end{figure}

\begin{itemize}
\item Unlike previous works (e.g. Boyle et al. 1994; Page et al. 1996;
  Jones et al. 1997),PLE was {absolutely rejected}. The 2D K-S 
  probabilities for the best-fit PLE models are less than $10^{-2}$ 
  for all sets of cosmological parameters considered 
  (see Table. \ref{tab:ldde_fit}). 
  
\item The best-fit PDE was marginally rejected with a 2D K-S probability of
  $\approx  5\%$). One difficulty of the best-fit PDE model is that it 
  overproduces the extragalactic soft CXRB intensity. 

\item The LDDE model, where the evolution rate drops at lower 
    luminosities, gave an acceptable overall 
    description of the SXLF (see below).

\end{itemize}

 The LDDE expression we have used for the overall SXLF is:
\begin{equation}
\frac{{\rm d}\;\Phi\,(L_{\rm x},z) }{{\rm d\;Log}\;L_{\rm x}}
  = {A}\;\left[(L_{\rm x}/{L_*})
      ^{{\gamma_1}}
         +(L_{\rm x}/{L_*})^{{\gamma_2}}
  	\right]^{-1}\cdot e(z,L_{\rm x}), 
\end{equation}
where $e(z,L_{\rm x})$ is the density evolution factor:
\begin{eqnarray}
	\left\{ 
	\begin{array}{ll}
	(1+z)^{\max(0,{p1}-{\alpha}({\rm Log}\; 
	  {L_{\rm a}} - {\rm Log}\;L_{\rm x}))} 
	     & (z \leq {z_{\rm c}}; L_{\rm x}<L_{\rm a}) \\ 
	(1+z)^{{p1}} 
	     & (z \leq {z_{\rm c}}; L_{\rm x}\ge L_{\rm a})\\ 
        e({z_{\rm c}},L_{\rm x})
         \left[(1+z)/(1+{z_{\rm c}}) \right]^{{p2}} 
	     & (z>{z_{\rm c}}) \\
	\end{array}
       \right. .\nonumber
\end{eqnarray}

 The parameter $\alpha$ represents the degree of 
luminosity dependence on the density evolution rate  for 
$L_{\rm x}<L_{\rm a}$. The PDE case is $\alpha=0$ and a greater 
value indicates lower evolution rates at low luminosities. 

 The best-fit parameters/90\% errors for the LDDE model, 2D K-S 
probabilities $P_{\rm 2DKS}$ (probabilities that the D-value for 
the 2D K-S statistics exceeds the observed value) are shown  
in Table \ref{tab:ldde_fit}. The integrated 0.5-2 keV intensity of the 
model, extrapolated  below the survey limit, is also shown.
This can be compared with the extragaxlactic CXRB intensity
of (7.4--9.0)$\times 10^{-12}[{\rm erg\,s^{-1}\,cm^{-2}\,deg^{-2}}]$,
from an update of the ROSAT/ASCA measurements by 
Miyaji et al. (1998a).

\begin{table}
\caption[]{The best-fit LDDE Parameters} 
\begin{center}
\begin{tabular}{ll}
\hline\hline
 $(\Omega_m,\Omega_\Lambda)$ & Parameters/2DKS probability/Intensity \\
\hline
(1.0,0.0) & $A=(1.57 \pm .11)\times 10^{-6}$;\,$L_*=0.57^{+.33}_{-.19}$ \\
     & $\gamma_1=0.68\pm .18$;\,$\gamma_2=2.26\pm .95$;\,$p1=5.4\pm .4$ \\
     & $z_{\rm c}=1.51\pm .15$;\,$p2=0.0$ (fixed) \\
     & $\alpha = 2.3\pm .8$;\, ${\rm Log} L_{\rm a}=44.2$ (fixed) \\
     & \underline{$P_{\rm 2DKS}$=51\%}; $I_{\rm x12}=4.5\pm 0.7$\\ 	
\\
(0.3,0.0) & $A=(2.18\pm .15)\times 10^{-6}$;\,$L_*=0.48^{+.21}_{-.13}$ \\
     & $\gamma_1=0.58\pm .20$;\, $\gamma_2=2.26\pm .08$;\, $p1=5.8\pm .4$ \\
     & $z_{\rm c}=1.54\pm .15$;\, $p2=0.0$ (fixed) \\
     & $\alpha = 2.6\pm .6$;\, ${\rm Log} L_{\rm a}=44.5$ (fixed) \\ 
     & \underline{$P_{\rm 2DKS}$=34\%}; $I_{\rm x12}=4.8\pm 0.7$\\ 	
\\
(0.3,0.7) & $A=(2.04\pm .14)\times 10^{-6}$;\,$L_*=0.50^{+.21}_{-.13}$ \\
     & $\gamma_1=0.58\pm .19$;\, $\gamma_2=2.27\pm .08$;\, $p1=5.8\pm .4$ \\
     & $z_{\rm c}=1.51\pm .14$;\, $p2=0.0$ (fixed) \\
     & $\alpha = 2.9\pm .6$;\, ${\rm Log} L_{\rm a}=44.6$ (fixed) \\ 
     & \underline{$P_{\rm 2DKS}$=38\%}; $I_{\rm x12}=4.6\pm 0.7$\\ 	
\\
(0.0,0.0) & $A=(2.35\pm .16)\times 10^{-6}$;\,$L_*=0.45^{+.19}_{-.12}$   \\
     & $\gamma_1=0.56\pm .20$;\, $\gamma_2=2.24\pm .08$;\, $p1=5.9\pm .4$   \\
     & $z_{\rm c}=1.55\pm .14$;\, $p2=0.0$ (fixed) \\
     & $\alpha = 2.7\pm .6$;\, ${\rm Log} L_{\rm a}=44.6$ (fixed) \\
     & \underline{$P_{\rm 2DKS}$=26\%}; $I_{\rm x12}=5.3\pm 0.9$ \\
	\\ 	
\hline
\end{tabular}
\end{center}
{\small Units -- A: [$h_{50}^3\;{\rm Mpc^{-3}}$],\,\,
 $L_*$: [$10^{44}\;h_{50}^{-2}{\rm erg\;s^{-1}}$],\,\,
$I_{\rm x12}$: $[10^{-12}{\rm erg\,s^{-1}\,cm^{-2}\,deg^{-2}}]$ 
in 0.5-2 keV}
\label{tab:ldde_fit} 
\end{table}

\begin{figure}[ht]
\resizebox{\hsize}{!}{\includegraphics{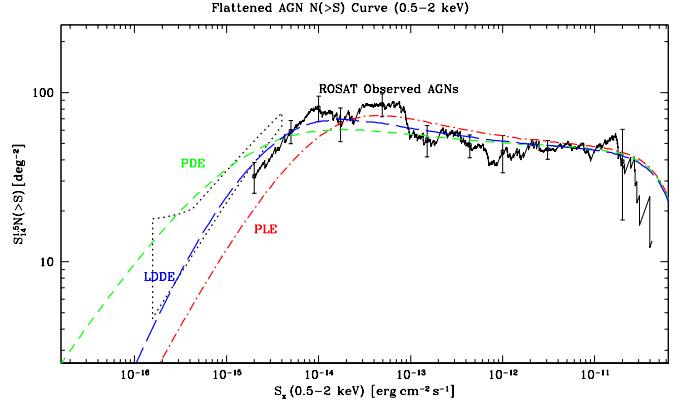}}
\caption[]{The $S^{1.5}N(>S)$ (a horizontal line corresponds to
the Eucleadean slope) curve for our sample AGNs is plotted with 90\% 
errors and are compared with the best-fit PLE (short-dashed), 
PDE (dot-dashed), and LDDE (dashed) models. 
The dotted fish is from the fluctuation analysis of the 
Lockman hole HRI data (including non-AGNs) by
Hasinger et al. (1998)}

\label{fig:fns}
\end{figure}

\begin{figure}[htbp]
\resizebox{\hsize}{!}{\includegraphics{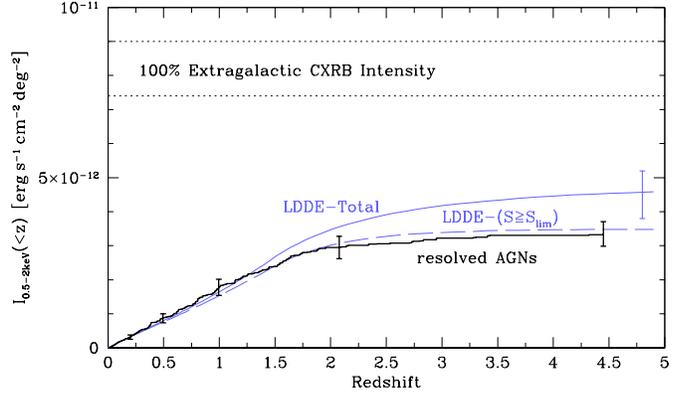}}
\caption[]{The cumulative intensity $I(<z)$ of the sample AGNs, defined 
by $\sum_{z_i<z} S_{{\rm x}i}/A(S_{{\rm x}i})$, where $S_{{\rm x}i}$ is
the flux of the object $i$ and $A(S_{{\rm x}i})$ is the available
survey area at this flux (Fig. \ref{fig:area}), is plotted as a function
of redshift (thick solid line with 90\% errors). The best-fit LDDE model 
is overplotted for the portion represented by the sample 
($S\geq S_{\rm lim}=1.9\times 10^{-15}$  $[{\rm erg\,s^{-1}\,cm^{-2}}]$; 
dashed line) and the total including
extrapolation to lower fluxes (thin solid line; with a 90\%
fitting error). Also 
the range of
the 0.5-2 keV extragalactic background intensity by  Miyaji et al.
1998a (updated) is shown.}
\label{fig:zi}
\end{figure}

 For demonstrations of the goodness of the overall fit, we 
compare the ``flattened''${\rm Log}\;N$ -- ${\rm Log}\;S$ 
($S^{1.5}N(>S)$) and cumulative $I(<z)$ curves of the real data 
with the model prediction in Figs. \ref{fig:fns} and \ref{fig:zi}
respectively. The results of the 2DKS test, along with these comparisons
show that our overall description is an excellent representation of the
current data.  The best-fit LDDE models, when
extrapolated below the survey limit, gives  50-70\% of the 
extragalactic background in the 0.5-2 keV band considering
errors of the fits. If the LDDE parameters 
are adjusted to produce 90\% of the extragalactic 0.5-2 keV background
(clusters should contribute $\sim 10\%$),  the 2D KS probability 
drops to $\sim 8\%$. 
 
\section{Evolution of QSO Number Density}

 Number density evolution of luminous QSOs is one of the important
pieces of key observational information on understanding  
blackhole formation and accretion history In particular, we
put emphasis on high luminosity QSOs (${\rm Log}\;L_{\rm}>44.5$),
where the XLF is consistent with the slope of $\gamma=2.3$ at all
redshift and free from complicated 
luminosity dependence of the evolution rate and contamination
from star formation activity.
 
\begin{figure}[t]
\begin{center}
\resizebox{0.8\hsize}{!}{\includegraphics{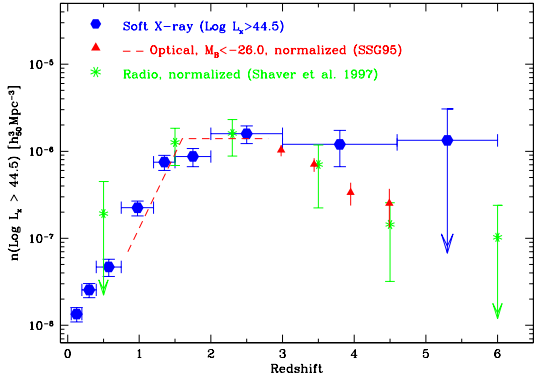}}
\end{center}
\caption[]{The comoving number density of luminous 
(${\rm Log}\;L_{\rm}>44.5$) QSOs in our ROSAT AGN sample are plotted 
as a function of redshift. For comparison, number density 
of optically-selected ($M_{\rm B}<-26$) (dashed line and filled
triangles, from SSG95) and radio-selected (stars, Shaver et al.
1997) QSOs, normalized to the soft X-ray selected QSO number density
at $z\sim 2.5$ are overplotted. Error bars are at the 90\% 
confidence level
and upper-limits correspond to 1 object per bin at the symbol and 90\% 
at the top.}
\label{fig:zev}
\end{figure}

 The comoving number density of the luminous QSOs in our sample
are plotted as a function of redshift in Fig. \ref{fig:zev}. 
For comparison, number densities (normalized to be approximately
equal to the soft X-ray point at $z\sim 2.5$) of optically-selected 
(Schmidt, Schneider, Gunn 1995, hereafter SSG95) and radio-selected 
(Shaver et al. 1997) QSOs  plotted as a function of redshift 
(see caption). 

 Unlike the optical and radio cases, we do not find evidence for
decrease of the space density beyond $z\approx 2.7$. Using the 
maximum-likelihood fitting, we have checked the significance 
of the inconsistency.  Requiring that the number density drops 
beyond $z=2.7$ as SSG95, the likelihood value (varies as $\chi^2$) 
increased by 3.3, showing that the significance of
the inconsistency is 93\%, which is marginal. In the $3.0<z\leq 4.6$,
we have 5 QSOs in the sample,  while expected number in the presence of
the decrease like the SSG95 result is 2.4. 
Including more surveys with a good completeness at the depth of  
$S_{\rm x}\approx 5\times 10^{-15}$  $[{\rm erg\,s^{-1}\,cm^{-2}}]$
would enable us to discuss whether the soft X-ray
selected QSO number density drops beyond $z\approx 3$.

\section{Discussion}
\label{sec:disc}

 There are some discrepancies between our work and previous
ones from several authors from combinations of the {\it Einsten} 
Extended Medium Sensitivity Survey (EMSS) and {\it ROSAT}
surveys (e.g. Boyle et al. 1994; Page et al. 1996; 
Jones et al. 1997), in that our results are not consistent with PLE. 
This is partially
due to the fact that our combined sample has more AGNs at fainter
fluxes, thus better statistics excludes simpler description. 
However, a direct comparision of the $1/V_{\rm a}$ estimates of 
the XLF between our sample and that of Jones et al., for example, 
shows no descrepancy in the $z<0.4$ bin, and high luminosity part 
(${\rm Log}\; L_{\rm x}\ga 44$) of the $z>0.4$, but descrepancies
appear at the low luminosity end of  the $z>0.4$ regime. The SXLFs
in the regime of bright sources, i.e. the RBS/SA-N based on the 
RASS for our work and the EMSS sample used by them, are mutually 
consistent. Thus the uncertainties in conversion of fluxes between 
{\it ROSAT} and {\it Einstein} bandpathes do not make significant 
contributions to the descrepancy. Two likely major causes
are: (1) inclusion of apparently narrow-line
objects which have indications of  AGN  activities in our work, while
they mainly argue the XLF for broad-line objects. (2): because of the
misidentification of the faintest X-ray sources ($S_{\rm x14}\la 0.5$) 
to field galaxies in their PSPC positioning, they miss some AGNs. A more 
complete analysis (e.g. comparing redshift distribution) using the original 
catalogs will be discussed in Miyaji et al. 1998b.  
   
 The extrapolations of our best-fit LDDE models explain 50-70\%
of the 0.5-2 keV extragalactic CXRB, considering errors of the fits. 
Clusters of galaxies are expected to contribute about 10\% in this band. 
The remaining contributors could be some low-luminosity galaxies/AGNs 
(${\rm Log}\;L_{\rm x}\la 10^{41.7}$) (see Fig. 2a of Hasinger 1998). 
These have about the same local volume emissivity as the AGNs 
in our SXLF, but uncertainties up to by a factor of few may exist, 
since this low-luminosity galaxy XLF is based on a sample of galaxies 
nearer than $7.15$ Mpc and the role of local overdensity can be important. 
Also practically no direct observational information exists for the 
evolution of these low luminosity sources.

 If the low-luminosity galaxies/AGNs contribute significantly to the
remaining CXRB, these can be:  (1): intrinsically low luminosity AGNs 
(2) high-intermediate redshift obscured AGNs, which are expected to 
contribute much of the harder CXRB and redshifted into the {\it ROSAT} band.
(3): star formation activity. 

 It is also possible that the behavior of the AGN SXLF 
(${\rm Log}\;L_{\rm x}\ga 10^{41.7}$) at intermediate-high 
redshifts does not follow the simple extrapolation of the current LDDE 
model below the luminosities corresponding to the survey limit at that
redshift. Especially our particular formula for the LDDE tends to
make the $N(<S)$ drop rapidly below the survey limit. Also the 
LDDE model-integrated intensity is sensitive to the lowest flux objects
in the sample, where incompleteness may play a role.  
These are fundamental uncertainties in the integrated intensity 
extrapolated from the best-fit representations of some functional form.  

\begin{acknowledgements}
Our work is deeply indebted to the effort of the observers and 
other members on the surveys used in our analysis. We particularly
thank the collaborators of the RBS, Marano, and RIXOS surveys
for allowing us to use the samples before publication. 
\end{acknowledgements}

\end{document}